\documentclass[aps,prd,twocolumn,amsmath,floats,floatfix,superscriptaddress,nofootinbib]{revtex4-2}
\usepackage[utf8]{inputenc}
\usepackage[english]{babel}
\usepackage{amsmath,amssymb,bm}
\usepackage{graphicx}
\usepackage{tikz}
\begin{document}
\title{Ernst-Potential Driven Recovery of Kerr with Neural Networks}
\author{Veselin G. Filev}
\email{vfilev@math.bas.bg}
\affiliation{Institute of Mathematics and Informatics, Bulgarian Academy of Sciences, Acad. G. Bonchev Str., 1113 Sofia, Bulgaria}
\author{Stoytcho S. Yazadjiev}
\affiliation{Institute of Mathematics and Informatics, Bulgarian Academy of Sciences, Acad. G. Bonchev Str., 1113 Sofia, Bulgaria}
\affiliation{Department of Theoretical Physics, Faculty of Physics, Sofia University, Sofia 1164, Bulgaria}

\begin{abstract}
We use artificial neural networks to recover the Kerr solution by directly minimizing a regularized Ernst sigma-model action. Reduction along the axial Killing direction avoids the divergence at the ergosphere. Compactified coordinates cover the entire black-hole exterior, while the network parametrization enforces the asymptotic data and axis regularity. For dimensionless spins $0.1 \le a/M \le 0.99$, the median maximum relative error in the recovered metric component $g_{\phi\phi}$ remains below $7.2\times10^{-4}$. When only the horizon rod length and angular momentum are prescribed, the networks also predict the mass with median relative errors below $1.8\times10^{-3}$. These results demonstrate the practicality of direct variational training for rotating black-hole geometries.
\end{abstract}

\maketitle

\section{Introduction}
\label{sec:introduction}
Physical reality is far too complex to be captured solely in terms of closed-form analytic solutions. In fact, the vast majority of real-world physical phenomena can be described quantitatively only with the aid of numerical methods.

It is therefore not surprising that computational approaches have long played an important role in theoretical physics, particularly in general relativity, where numerical relativity has become an essential tool for understanding fascinating phenomena such as black-hole mergers~\cite{Pretorius:2005gq,Campanelli:2005dd,Baker:2005vv}, neutron-star collisions~\cite{Shibata:1999wm}, and the gravitational waves they produce~\cite{LIGOScientific:2016aoc,LIGOScientific:2017vwq}.

But what does it mean to have a numerical description? It might, for example, take the form of a spline approximation to the solution of a differential equation or a Monte Carlo simulation of a physical process. Alternatively, the universal approximation theorem for artificial neural networks (ANNs) guarantees that ANNs can approximate any continuous function (on a compact domain) to arbitrary accuracy~\cite{Cybenko:1989,Hornik:1989}. We should therefore be able to capture physical reality equally well in terms of a neural network. 

One of the main approaches to using ANNs to solve physics problems governed by equations of motion (EOMs) is to minimize the residual of the EOMs. This approach, often combined with an architecture realizing the symmetries of the problem, is known as Physics-Informed Neural Networks (PINNs)~\cite{Raissi:2019}. It is a powerful approach which has led to numerous remarkable contributions.

However, in theories with well-defined least-action variational principles, the EOMs are precisely the Euler--Lagrange equations extremizing the action functional. As such, the resulting differential equations contain higher-order derivatives than the action itself, which requires a higher working precision to obtain the desired numerical accuracy.

On the other hand, the training of a neural network is, in essence, a global minimization task. Thus, if we manage to formulate the least-action principle as a minimization problem for a properly regularized action (e.g., free energy), we could train the neural network by directly minimizing this action, avoiding the step of deriving complicated EOMs and reducing the order of the required derivatives.

Recently, this approach has been successfully applied in the context of holographic flavor~\cite{Filev:2025mbt} and holographic entanglement entropy~\cite{Filev:2026gka}, where ANNs were trained to directly minimize a regularized DBI action and the area functional of Ryu--Takayanagi surfaces. In this work, our goal is to apply this approach in the context of general relativity, in particular to the numerical recovery of the Kerr metric~\cite{Kerr:1963}.

A natural starting point would be to consider the regularized, Wick-rotated Einstein--Hilbert action, that is, the free energy. Unfortunately, most asymptotically flat solutions to general relativity are actually saddle points~\cite{Gross:1982} which do not minimize the off-shell free energy functional. This leads to escape along flat directions during the training process. The free energy could still be used when combined with min-max optimization, but this complicates the training procedure and it is no longer competitive with minimizing the residuals (e.g., the PINN approach).

Fortunately, it is still possible to reformulate the variational problem as a minimization one by using a properly defined Ernst potential. In the next section we will review the Ernst sigma model in Weyl–Papapetrou coordinates.

\section{The Conjugate Ernst Sigma Model}
\label{sec:Ernst}
A stationary axisymmetric vacuum spacetime admits two Killing vectors $\partial_t$ and $\partial_\phi$ and can be put in Weyl–Papapetrou form:
\begin{equation}
\begin{split}
    ds^2 = {}&-f(\rho,z)\bigl(dt - \omega(\rho,z)\,d\phi\bigr)^2 \\
         &+ f(\rho,z)^{-1}\Bigl[ e^{2\gamma(\rho,z)}\bigl(d\rho^2 + dz^2\bigr) \\
         &\hphantom{{}+ f(\rho,z)^{-1}\Bigl[} + \rho^2\,d\phi^2 \Bigr].
\end{split}
    \label{eq:wp_metric}
\end{equation}
Defining the twist potential $b$ via
\begin{equation}
\nabla_a b = -\rho^{-1}\,f^2\,\epsilon_{ab}\nabla^b\omega ,
    \qquad \epsilon_{\rho z} = +1 ,
    \label{eq:std_twist}
\end{equation}
one can show that the sigma model action
\begin{equation}
    S_{\rm std} = \int_{\rho>0} \rho\,
       \frac{|\nabla f|^2 + |\nabla b|^2}{f^2}\,d\rho\,dz,
    \label{eq:S_standard}
\end{equation}
has the vacuum Einstein equations as its Euler--Lagrange equations, which in terms of $f$ and $b$ are equivalent to the Ernst equations~\cite{Ernst:1968}:
\begin{equation}
     f\,\Delta f = |\nabla f|^2 - |\nabla b|^2 ,
     \qquad
     f\,\Delta b = 2\,\nabla f\!\cdot\!\nabla b ,
     \label{eq:std_ernst}
\end{equation}
where $\nabla = (\partial_\rho,\partial_z)$ and $\Delta = \partial_\rho^2 + \rho^{-1}\partial_\rho + \partial_z^2$ is the flat Laplacian in cylindrical coordinates. The equations can also be written in terms of the complex Ernst potential $\mathcal{E} = f + i\,b$:
\begin{equation}
(\operatorname{Re}\mathcal{E})\,\Delta\mathcal{E}
      = \nabla\mathcal{E}\cdot\nabla\mathcal{E} .
     \label{eq:ernst_complex}
 \end{equation}
Note that the ansatz (\ref{eq:wp_metric}) can be thought of as a Kaluza–Klein reduction along the stationary Killing direction $\partial_t$, while $b$ is a twist potential dual to the associated fiber connection $\omega$. This suggests that a similar construction should be possible along the Killing direction $\partial_\phi$. However, although positive, the density of (\ref{eq:S_standard}) diverges in the bulk of the geometry when crossing an ergosphere. At the ergosphere the norm of the Killing vector $\partial_t$ vanishes (that is, $g_{tt} = -f = 0$). In principle, this could be regulated; however, the fact that we do not know the position of the ergosphere complicates the procedure. Moreover, the configuration space is split into two disconnected regions, outside ($f>0$) and inside ($f<0$) the ergosphere, which further complicates the set-up. 

Fortunately, the norm of the Killing vector $\partial_\phi$ does not vanish at the ergosphere, which is why we reduce along $\partial_\phi$ instead of $\partial_t$ to obtain the so-called conjugate (Matzner--Misner) form of the metric~\cite{Matzner:1967}:
\begin{equation}
\begin{split}
    ds^2 = {}& F(\rho,z)\bigl(d\phi + \Omega(\rho,z)\,dt\bigr)^2 \\
         &+ F(\rho,z)^{-1}\Bigl[ e^{2\hat\gamma(\rho,z)}\bigl(d\rho^2 + dz^2\bigr)
           - \rho^2\,dt^2 \Bigr],
\end{split}
    \label{eq:conj_metric}
\end{equation}
where $F = g_{\phi\phi}$ is the norm of the Killing vector $\partial_\phi$ and $\Omega = g_{t\phi}/g_{\phi\phi}$. Note that even though the $(t, \phi)$
 block of the metric is parameterized differently in (\ref{eq:wp_metric}) and (\ref{eq:conj_metric}), they still share the same Weyl coordinate $\rho$, defined as the volume of the internal fiber: $\rho^2 = g_{t\phi}^2 - g_{tt}\,g_{\phi\phi}$. Note that the real part of the standard Ernst potential can be obtained algebraically in terms of $F$ and $\Omega$ as:
\begin{equation}
    f = \frac{\rho^2}{F} - F\,\Omega^2 .
    \label{eq:f_from_F}
\end{equation}
Equation (\ref{eq:f_from_F}) expresses $f$ as a competition of the squared lapse of the zero-angular-momentum observers $N^2=\rho^2/F$ and the spacelike rotational term $F\,\Omega^2$, with the lapse dominating outside the ergosphere and the rotational term dominating inside it. On the other hand, $F$ is positive everywhere off the axis of symmetry, including at the ergosphere and up to the horizon. Therefore, the conjugate fields are regular at the ergosphere, whose location is determined from (\ref{eq:f_from_F}).

\subsection{The conjugate Ernst equations}
\label{sec:conj_ernst}

The conjugate twist potential $B$ dual to the fiber connection $\Omega$ is defined via:
\begin{equation}
    \nabla_a B = -\rho^{-1}\,F^2\,\epsilon_{ab}\nabla^b\Omega ,
    \qquad \epsilon_{\rho z} = +1 .
    \label{eq:conj_twist}
\end{equation}
Note that the metric (\ref{eq:conj_metric}) can be obtained from the metric (\ref{eq:wp_metric}) via the formal substitution: $t \to i\phi$, $\phi \to i t$, with $F = f$ and $\Omega = -\omega$. Therefore, equation (\ref{eq:conj_twist}) suggests the map $B=-b$. Correspondingly, the standard Ernst equations transform into: 
\begin{equation}
    F\,\Delta F = |\nabla F|^2 - |\nabla B|^2 ,
    \qquad
    F\,\Delta B = 2\,\nabla F\!\cdot\!\nabla B .
    \label{eq:conj_ernst}
\end{equation}
Equations (\ref{eq:conj_ernst}) are precisely the Euler--Lagrange equations of the conjugate sigma model:
\begin{equation}
    S_{\rm conj} = \int_{\rho>0} \rho\,
       \frac{|\nabla F|^2 + |\nabla B|^2}{F^2}\,d\rho\,dz \ .
    \label{eq:S_conj}
\end{equation}
In contrast to (\ref{eq:S_standard}), the density of (\ref{eq:S_conj}) is finite away from the axis $\rho=0$, including inside the ergosphere. It diverges only on the part of the axis where $F$ vanishes as $F\sim \rho^2$. Given that $F$ is strictly positive off the axis, the pair $(F,B)$ plays the role of coordinates on the upper half-plane. Next, we remove the divergence at the axis.

% ---------------------------------------------------------------------------
% Continuation of ernst_chapter.tex. Drafted by Claude (AI) on 2026-10-18 and
% rewritten in the author's style; to be edited by the author.
% Conventions: y is the CONVENTIONAL prolate variable (y -> cos(theta) at large
% r, y = +1 on the upper outer axis). The code (prolate_nn/) uses y_code = -y;
% every formula here was checked numerically against the code's exact Kerr.
% Citation placeholders follow the style of introduction.tex.
% ---------------------------------------------------------------------------

\subsection{Axis behavior and the reference decomposition}
\label{sec:reference}
Let us consider a Kerr black hole of mass $M$ and angular momentum $J$\footnote{$M$ and $J$ are defined by the asymptotic behavior $g_{tt}\to -(1-2M/r)$ and $g_{t\phi}\to -2J\,\sin^2\theta/r$ at large $r$, with $r = \sqrt{\rho^2 + z^2}$ and $\cos\theta = z/r$.}. In Weyl coordinates the $\rho=0$ axis has two parts: the ``rod'' at $|z| <\sigma$ (here $\sigma = \sqrt{M^2-a^2}$ and $a=J/M$) corresponding to the horizon, where $F$ remains finite, and the region $|z| > \sigma$, where $F \sim \rho^2$. The latter is the source of the divergence in (\ref{eq:S_conj}). Note that this behavior is universal for any regular black hole with a single horizon. In particular, the Schwarzschild solution with mass $\sigma$ has the same rod, while $F = \rho^2\,e^{U_{\rm ref}}$ with:
\begin{equation}
    U_{\rm ref} = \ln\frac{R_+ + R_- + 2\sigma}{R_+ + R_- - 2\sigma},
    \qquad
    R_\pm = \sqrt{\rho^2 + (z\mp\sigma)^2} ,
    \label{eq:Uref}
\end{equation}
which is harmonic away from the rod, $\Delta U_{\rm ref} = 0$, and behaves as $-2\ln\rho$ on it. Therefore, we can factor out the singular behavior of $F$ by writing:
\begin{equation}
    F = \rho^2\, e^{\,U_{\rm ref} + U} ,
    \label{eq:F_split}
\end{equation}
and have the neural network learn only the remaining field $U$, which is smooth and bounded everywhere, including the horizon and the axis. 

Note that for $a = 0$, when $\sigma = M$ and Schwarzschild is the exact solution, we have $U=0$ and $B=0$. Therefore, $U$ and $B$ are non-trivial only for $a > 0$ ($\sigma < M$), encoding everything that the rotation adds.

After substituting (\ref{eq:F_split}) into (\ref{eq:S_conj}) and ignoring total derivatives, we obtain the regularized action:
\begin{equation}
    E[U,B] = \int_{\rho>0} \rho\Bigl[\,|\nabla U|^2
       + \frac{|\nabla B|^2}{F^2}\Bigr]\,d\rho\,dz .
    \label{eq:E_def}
\end{equation}
Using that $\Delta\ln\rho = 0$ and $\Delta U_{\rm ref} = 0$, one can show that the Euler--Lagrange equations of (\ref{eq:E_def})
\begin{equation}
    \Delta U + \frac{|\nabla B|^2}{F^2} = 0 ,
    \qquad
    \nabla\!\cdot\!\Bigl(\frac{\rho}{F^2}\,\nabla B\Bigr) = 0 ,
    \label{eq:EL}
\end{equation}
are exactly the conjugate Ernst equations (\ref{eq:conj_ernst}). Note that (\ref{eq:E_def}) is manifestly positive, which makes it a natural candidate for a loss function. Whether the dropped total derivatives matter depends on the boundary conditions imposed on $U$ and $B$. In Sec.~\ref{sec:class} we show that for our choice of boundary conditions all boundary contributions to the first variation of $E$ vanish identically. Therefore, the exact solution is a stationary point of (\ref{eq:E_def}) without adding any boundary terms.

\subsection{Compactified prolate coordinates}
\label{sec:prolate}

In order to work with the full exterior of the black hole, it is convenient to introduce prolate spheroidal coordinates adapted to the rod:
\begin{equation}
    x = \frac{R_+ + R_-}{2\sigma} ,
    \qquad
    y = \frac{R_- - R_+}{2\sigma} ,
    \label{eq:prolate}
\end{equation}
in terms of which $\rho = \sigma\sqrt{(x^2-1)(1-y^2)}$ and $z = \sigma x y$. The horizon rod is at $x = 1$, while the upper and lower components of the axis are at $y = +1$ and $y = -1$, respectively. At large distances $x \simeq r/\sigma$ and $y \to \cos\theta$. Note that the flat metric is diagonal in these coordinates:
\begin{equation}
    d\rho^2 + dz^2 = \sigma^2(x^2 - y^2)
    \Bigl[\frac{dx^2}{x^2-1} + \frac{dy^2}{1-y^2}\Bigr] ,
\end{equation}
and when combined with the measure of (\ref{eq:E_def}) it gives:
\begin{equation}
\begin{split}
    \rho\,\nabla u\!\cdot\!\nabla v\;d\rho\,dz
    = {}&\sigma\bigl[(x^2-1)\,u_x v_x \\
        &+ (1-y^2)\,u_y v_y\bigr]\,dx\,dy ,
\end{split}
    \label{eq:measure}
\end{equation}
which is free of singular factors. In these coordinates the reference potential takes the simple form $U_{\rm ref} = \ln[(x+1)/(x-1)]$, and $F = \sigma^2(x+1)^2(1-y^2)\,e^{U}$.

\begin{figure}[t]
\centering
\begin{tikzpicture}[font=\footnotesize]
\colorlet{hor}{red!75!black}
\colorlet{ax}{blue!70!black}
% ---- (a) Weyl half-plane ----
\begin{scope}[scale=0.85]
  \def\R{2.4}
  \begin{scope}
    \clip (0,-\R) -- (0,\R) arc[start angle=90, end angle=-90, radius=\R] -- cycle;
    \foreach \xx in {1.25,1.6667,2.2222}
      \draw[gray!55] plot[domain=-90:90, samples=61, variable=\u, smooth]
        ({sqrt(\xx*\xx-1)*cos(\u)}, {\xx*sin(\u)});
    \foreach \yy in {-0.5,0,0.5}
      \draw[gray!55] plot[domain=0:2.2, samples=41, variable=\v, smooth]
        ({sinh(\v)*sqrt(1-\yy*\yy)}, {cosh(\v)*\yy});
  \end{scope}
  \draw[gray, dashed] (0,\R) arc[start angle=90, end angle=-90, radius=\R];
  \draw[ax, very thick] (0,1) -- (0,\R);
  \draw[ax, very thick] (0,-1) -- (0,-\R);
  \draw[hor, very thick] (0,-1) -- (0,1);
  \fill (0,1) circle (1.6pt);
  \fill (0,-1) circle (1.6pt);
  \node[hor, rotate=90, above] at (0,0) {horizon};
  \node[ax, rotate=90, above] at (0,1.7) {axis};
  \node[ax, rotate=90, above] at (0,-1.7) {axis};
  \node[above] at (0,\R) {$z$};
  \node[right] at (\R,0) {$\rho$};
  \node[gray, above right] at (45:\R) {$\infty$};
  \node at (-0.75,\R+0.25) {(a)};
\end{scope}
% ---- (b) compactified rectangle ----
\begin{scope}[xshift=3.9cm, x=2.6cm, y=2.04cm]
  \foreach \ss in {0.45,0.6,0.8} \draw[gray!55] (\ss,-1) -- (\ss,1);
  \foreach \yy in {-0.5,0,0.5} \draw[gray!55] (0,\yy) -- (1,\yy);
  \draw[gray, dashed] (0,-1) -- (0,1);
  \draw[ax, very thick] (0,1) -- (1,1);
  \draw[ax, very thick] (0,-1) -- (1,-1);
  \draw[hor, very thick] (1,-1) -- (1,1);
  \fill (1,1) circle (1.6pt);
  \fill (1,-1) circle (1.6pt);
  \node[ax, above] at (0.5,1) {axis, $y=+1$};
  \node[ax, below] at (0.5,-1) {axis, $y=-1$};
  \node[hor, rotate=-90, above] at (1,0) {horizon, $s=1$};
  \node[gray, rotate=90, above] at (0,0) {infinity, $s=0$};
  \node at (-0.25,1.18) {(b)};
\end{scope}
\end{tikzpicture}
\caption{The exterior of the black hole in Weyl coordinates (a) and in the compactified prolate coordinates $(s,y)$ (b), for $\sigma = 1$. The horizon rod (red) is mapped to the edge $s = 1$, the upper and lower components of the axis (blue) to $y = \pm1$, and spatial infinity (dashed) to $s = 0$. The endpoints of the rod, $z = \pm\sigma$ (dots), become the corners $(1,\pm1)$. The thin lines are the coordinate lines $s = 0.45, 0.6, 0.8$ (confocal half-ellipses in (a)) and $y = 0, \pm0.5$ (hyperbolae in (a)).}
\label{fig:domain}
\end{figure}
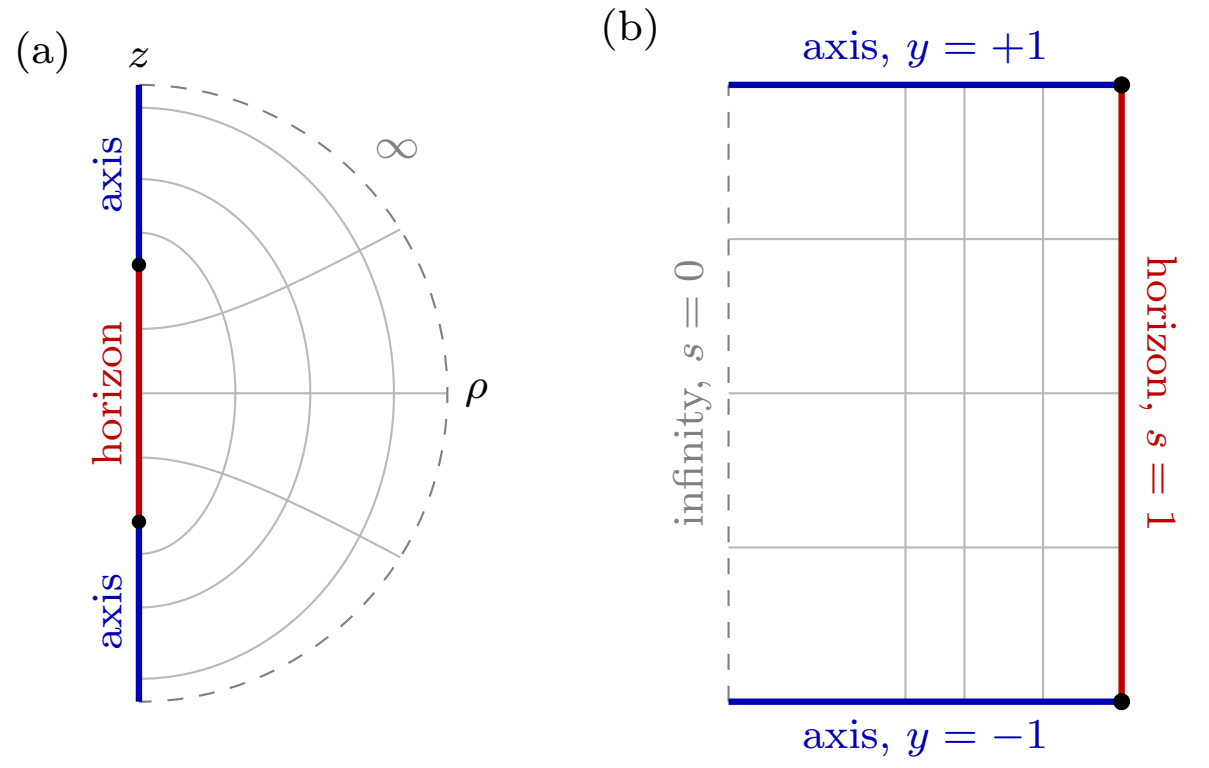

Finally, we compactify the radial direction by introducing $s = 1/x \in (0,1]$, which maps the entire exterior of the black hole onto the rectangle $(s,y)\in(0,1]\times[-1,1]$, with spatial infinity at $s = 0$ and the horizon at $s = 1$ (see Fig.~\ref{fig:domain}).
% Sketch of the integration domain, drafted by Claude (AI) on 2026-10-18.
% (a) Weyl half-plane (rho, z) for sigma = 1; (b) compactified rectangle (s, y).
% Coordinate lines: s = 0.45, 0.6, 0.8 (x = 1/s) and y = 0, +-0.5.

 Using (\ref{eq:measure}), $\partial_x = -s^2\partial_s$ and $dx = -ds/s^2$, the action (\ref{eq:E_def}) becomes:
\begin{equation}
\begin{split}
    E = \sigma\!\int_0^1\!\!ds\!\int_{-1}^{1}\!\!dy\,
    \Bigl[ &(1-s^2)\,U_s^2 + \frac{1-y^2}{s^2}\,U_y^2 \\
    &+ \frac{(1-s^2)\,B_s^2 + (1-y^2)\,B_y^2/s^2}{F^2} \Bigr] ,
\end{split}
    \label{eq:E_rect}
\end{equation}
where $F = \sigma^2(1+s)^2(1-y^2)\,e^{U}/s^2$. Note that no truncation of the domain is needed, since the integral extends over the full exterior.

\subsection{Boundary data and the admissible class}
\label{sec:class}

The mass and the angular momentum of the black hole enter only through the behavior of the fields at infinity and on the axis. At large $r$ the field $U_{\rm ref} + U = \ln(F/\rho^2)$ is asymptotically harmonic, and its monopole is fixed by the ADM mass: $U_{\rm ref} + U = 2M/r + O(r^{-2})$. Since $U_{\rm ref} \to 2\sigma/r$, we have:
\begin{equation}
    U = c_m\, s + O(s^2) ,
    \qquad
    c_m = 2\Bigl(\frac{M}{\sigma} - 1\Bigr) .
    \label{eq:U_asym}
\end{equation}
On the other hand, the frame dragging $g_{t\phi} \to -2J\sin^2\theta/r$ gives $\Omega \to -2J/r^3$ and, through (\ref{eq:conj_twist}), a twist potential with an $r$-independent angular profile:
\begin{equation}
    B \to B_\infty(y) = -2J\,y\,(3 - y^2) ,
    \label{eq:Binf}
\end{equation}
which takes the values $B = \mp4J$ on the upper and lower components of the axis. Note that (\ref{eq:Binf}) is not the dipole $-4J\cos\theta$, since the difference $-2Jy(1-y^2)$ does not decay with distance. Regularity on the axis further requires $\partial_\rho B = O(\rho^3)$, which is satisfied by (\ref{eq:Binf}).

We impose all of these conditions exactly by writing:
\begin{align}
    U &= c_m\, s + s^2\, N(s,y) , \label{eq:U_class}\\
    B &= B_\infty(y) + (1-y^2)^2\, s^2\, G(s,y) , \label{eq:B_class}
\end{align}
where $N$ and $G$ are smooth functions on the closed rectangle, with $N$ even and $G$ odd in $y$ (reflection symmetry $z \to -z$). The factor $s^2$ leaves the monopole (\ref{eq:U_asym}) and the profile (\ref{eq:Binf}) intact, while the factor $(1-y^2)^2$ preserves the axis values $\mp4J$ and the $O(\rho^3)$ behavior of $\partial_\rho B$. Note that with (\ref{eq:U_class})--(\ref{eq:B_class}) the singular factors $1/s^2$ and $(1-y^2)^{-2}$ in (\ref{eq:E_rect}) cancel, and the integrand is bounded on the closed rectangle, including the horizon, the axis and spatial infinity (see Appendix~\ref{app:class}).

Furthermore, all boundary terms in the first variation of (\ref{eq:E_rect}) vanish identically for admissible variations (see Appendix~\ref{app:class}). Therefore, we do not need to add any boundary terms to (\ref{eq:E_rect}), and every solution of (\ref{eq:EL}) in the admissible class is a stationary point of $E$.

Furthermore, if we treat the coefficient $c_m$ in (\ref{eq:U_class}) as unknown, only $\sigma$ and $J$ are prescribed, and the mass becomes a prediction:
\begin{equation}
    M = \sigma\Bigl(1 + \frac{c_m}{2}\Bigr) .
    \label{eq:M_pred}
\end{equation}
Since the corresponding variation $\delta U = s\,\delta c_m$ also vanishes at infinity, the exact solution remains a stationary point in this enlarged class.

\subsection{The Kerr solution in these variables}
\label{sec:kerr}

Finally, let us write the Kerr solution in these variables. Defining $p = \sigma/M$ and $q = a/M$ (so that $p^2 + q^2 = 1$), together with:
\begin{align}
    \mathcal{D} &= (1 + p x)^2 + q^2 y^2 , \nonumber\\
    \mathcal{A} &= \bigl[(1 + p x)^2 + q^2\bigr]^2 - p^2 q^2 (x^2 - 1)(1 - y^2) ,
\end{align}
the Kerr solution takes the form:
\begin{equation}
    F = M^2 (1 - y^2)\,\frac{\mathcal{A}}{\mathcal{D}} ,
    \qquad
    B = B_\infty(y) - \frac{2Jq^2\, y\,(1 - y^2)^2}{\mathcal{D}} .
    \label{eq:kerr}
\end{equation}
In terms of (\ref{eq:U_class})--(\ref{eq:B_class}) this corresponds to:
\begin{equation}
    U = \ln\frac{M^2 \hat{\mathcal{A}}}{\sigma^2 (1+s)^2\,\hat{\mathcal{D}}} ,
    \qquad
    G = -\frac{2Jq^2\, y}{\hat{\mathcal{D}}} ,
    \label{eq:kerr_class}
\end{equation}
where $\hat{\mathcal{D}} = s^2\mathcal{D} = (s + p)^2 + q^2 y^2 s^2$ and $\hat{\mathcal{A}} = s^4\mathcal{A} = \bigl[(s + p)^2 + q^2 s^2\bigr]^2 - p^2 q^2 s^2 (1 - s^2)(1 - y^2)$. Since $\hat{\mathcal{D}} \geq p^2 > 0$, both $G$ and $N = (U - c_m s)/s^2$ are smooth and bounded on the closed rectangle for every $a < M$, and therefore the Kerr solution belongs to the admissible class. Evaluating (\ref{eq:E_rect}) on (\ref{eq:kerr_class}) numerically gives $E = 0.630850791$ for $a = M/2$ (converged to all digits shown), while its first variation along generic admissible directions vanishes to machine precision. % [author: keep here or move to Sec. results]
% Claude draft, continues the last sentence

\section{Neural network and training}
\label{sec:training}
% ---------------------------------------------------------------------------
% Sec. "Neural network and training". Drafted by Claude (AI) on 2026-10-26 in the
% style of the author's sections; to be edited by the author. The numbers are
% those of the code (prolate_nn/model.py, train.py) and of the runs on the DIAS
% node c-18 (copied to prolate_nn/c18). References are in references.bib.
% ---------------------------------------------------------------------------

\begin{figure*}[t]
\centering
\begin{tikzpicture}[font=\small, >=stealth]
  % ---- inputs, shared by both networks ----
  \node[circle,draw,minimum size=1.3cm,inner sep=1pt] (I1) at (0,0.8) {$2s-1$};
  \node[circle,draw,minimum size=1.3cm,inner sep=1pt] (I2) at (0,-0.8) {$2y^2-1$};
  % ---- the two networks: hidden layers ----
  \foreach \b/\yc/\out in {T/2.15/N, B/-2.15/g} {
    \foreach \l/\x/\col in {1/2.5/blue!10, 2/4.8/green!10, 3/7.1/blue!10} {
      \foreach \i in {1,...,5} {
        \node[circle,draw,fill=\col,minimum size=0.5cm] (H\b\l\i) at (\x,\yc+1.16-0.58*\i+0.58) {};
      }
    }
    \node[circle,draw,fill=red!10,minimum size=0.95cm] (O\b) at (9.1,\yc) {$\out$};
    % connections
    \foreach \i in {1,...,5} {
      \draw[->,thick,opacity=0.55] (I1) -- (H\b1\i);
      \draw[->,thick,opacity=0.55] (I2) -- (H\b1\i);
      \draw[->,thick] (H\b3\i) -- (O\b);
      \foreach \j in {1,...,5} {
        \draw[->,opacity=0.3] (H\b1\i) -- (H\b2\j);
        \draw[->,opacity=0.3] (H\b2\i) -- (H\b3\j);
      }
    }
  }
  % layer labels, as in the reference figure
  \foreach \l in {1,2,3} {
    \node[above=1pt,align=center,font=\scriptsize] at (HT\l1.north) {Linear\\ + Tanh};
    \node[below=1pt,align=center,font=\scriptsize] at (HB\l5.south) {Linear\\ + Tanh};
  }
  \node[above=1pt] at (OT.north) {\scriptsize Linear};
  \node[below=1pt] at (OB.south) {\scriptsize Linear};
  % ---- admissible class and loss ----
  \node[draw,rounded corners=3pt,fill=gray!8,inner sep=5pt] (U) at (12.3,2.15) {$U = c_m\, s + s^2 N$};
  \node[draw,rounded corners=3pt,fill=gray!8,inner sep=5pt] (B) at (12.3,-2.15) {$B = B_\infty(y) + (1-y^2)^2 s^2\, y\, g$};
  \node[draw,rounded corners=3pt,fill=orange!15,inner sep=6pt,align=center] (E) at (16.2,0) {loss\\[2pt] $E[U,B]$};
  \draw[->,thick] (OT) -- (U);
  \draw[->,thick] (OB) -- (B);
  \draw[->,thick] (U.east) -- (E.north west);
  \draw[->,thick] (B.east) -- (E.south west);
\end{tikzpicture}
\caption{Architecture of the two neural networks used to learn the functions $N$ and $G = y\,g$ of the admissible class (\ref{eq:U_class})--(\ref{eq:B_class}). Both networks take $(2s-1,\,2y^2-1)$ as an input and have three hidden layers; for visual simplicity only five neurons per hidden layer are shown, while the networks used in practice have $32$. The outputs give the fields $U$ and $B$, and the action (\ref{eq:E_rect}) of these fields is the loss function. In the free-mass mode $c_m$ is a trainable parameter as well.}
\label{fig:ANN}
\end{figure*}
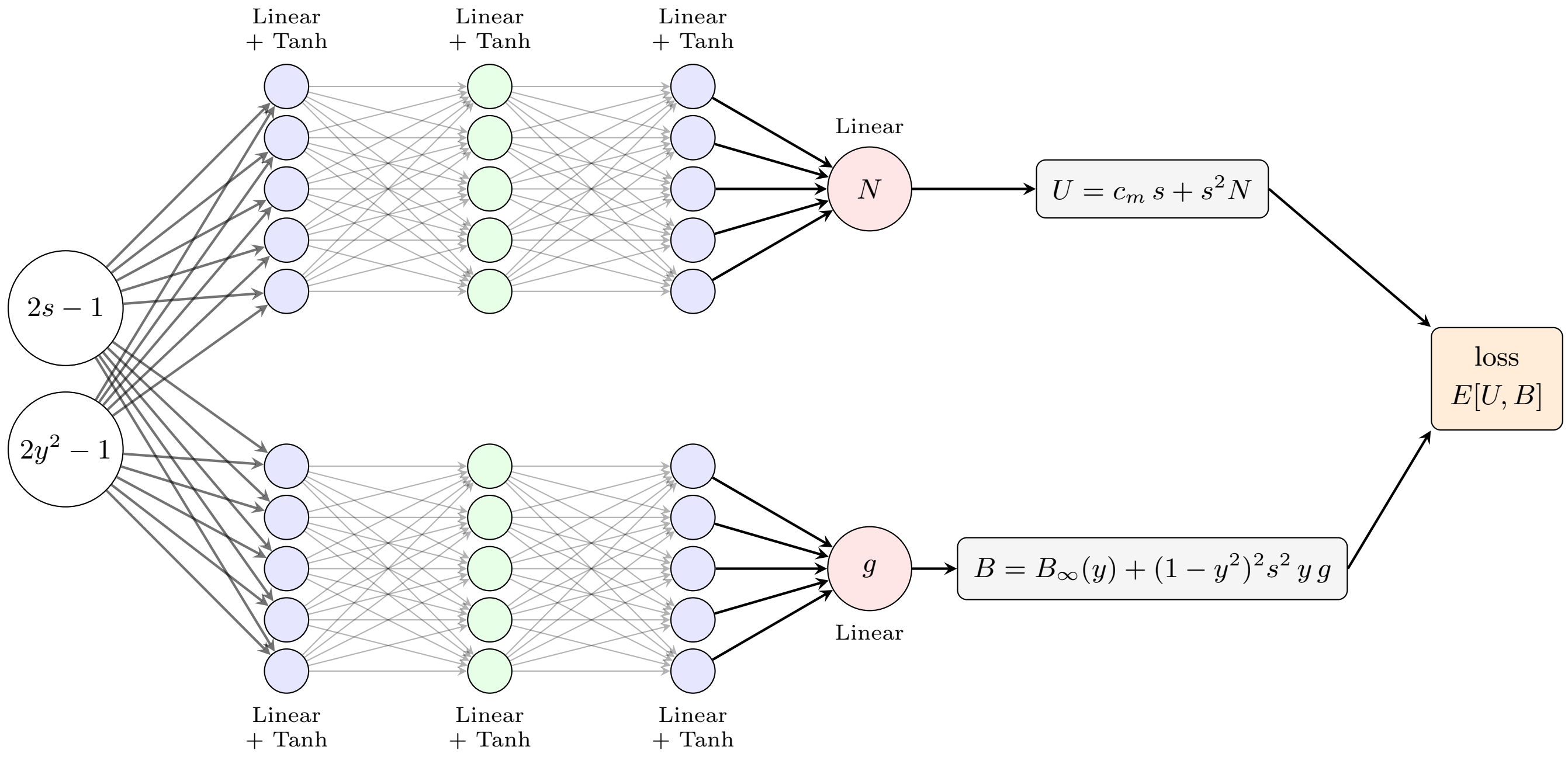

The admissible class (\ref{eq:U_class})--(\ref{eq:B_class}) leaves us with two unknown functions on the rectangle, $N(s,y)$ and $G(s,y)$. We represent each of them by a fully connected neural network with three hidden layers of $32$ neurons and $\tanh$ activation (see Fig.~\ref{fig:ANN}). Both networks take as an input the pair $(2s-1,\,2y^2-1)$, and we write $G = y\,g$, where $g$ is the output of the second network. Therefore, $N$ is even and $G$ is odd in $y$, and the reflection symmetry $z \to -z$ is built in exactly. Together the two networks have $4482$ parameters.

% Network architecture, in the style of Fig. 1 of arXiv:2506.20115 (Holographic flavour and neural
% networks). Drafted by Claude (AI) on 2026-10-26. Two networks, for N and for g (G = y g), with three
% hidden layers each (32 tanh neurons in practice, five drawn); their outputs enter the admissible class
% (eq:U_class)-(eq:B_class) and the action (eq:E_rect) of the resulting fields is the loss.

Note that no basis functions or any other information about the Kerr solution enter this representation. The behavior at infinity, on the axis and on the horizon is carried by the explicit factors in (\ref{eq:U_class})--(\ref{eq:B_class}), while everything else is learned. We initialize the weights of the output layers close to zero, so that the training starts from $U = c_m s$ and $B = B_\infty(y)$, that is, from the monopole and the asymptotic twist alone.

\subsection{The loss function and the optimization}
\label{sec:loss}

The loss function is the action (\ref{eq:E_rect}) itself, and no other terms are added to it. Since its integrand is bounded on the closed rectangle, it can be evaluated by standard numerical quadrature, while the derivatives $U_s$, $U_y$, $B_s$ and $B_y$ are obtained by automatic differentiation of the networks. Note that only first derivatives of the networks enter the loss, while the residuals of the equations of motion (\ref{eq:EL}) would require second derivatives. All computations are done in double precision using PyTorch~\cite{Paszke:2019xhz}.

The training proceeds in two stages. In the first stage we use the Adam optimizer~\cite{Kingma:2014vow} for $6000$ steps, with a learning rate decreasing from $2\times10^{-3}$ to $10^{-5}$. At every step the action is estimated on a new set of random points, one point in each cell of a uniform $64\times64$ partition of the rectangle. Since the points change at every step, the estimate is unbiased and the networks cannot adapt to any particular set of points.

In the second stage we refine the solution with the quasi-Newton L-BFGS method~\cite{Liu:1989}, which requires a fixed loss function. Therefore, we evaluate the action on a fixed composite Gauss--Legendre rule with $64$ nodes in each direction, and perform $300$ steps of at most $50$ iterations each.

Training on a fixed set of nodes, however, has a pitfall: the optimizer can lower the quadrature estimate of the action rather than the action itself, by developing oscillations between the nodes. We observed this in a few of our runs, in which the estimate fell below the Kerr value of the action, which is impossible for the exact integral. To prevent this, after every L-BFGS step we also evaluate the action on a second, finer Gauss--Legendre rule with $192$ nodes in each direction, and accept the step only if this value did not increase. Otherwise the step is undone and the training continues on a finer training rule. Note that this check does not involve the exact solution, and as long as it accepts every step the training is unaffected by it. In all runs reported in Sec.~\ref{sec:results} the two rules agree to better than $10^{-10}$ at the end of the training.

\subsection{Fixed and free mass}
\label{sec:modes}

We train the networks in two modes. In the first one the mass is fixed and $c_m$ is given by (\ref{eq:U_asym}). In the second one $c_m$ is an additional trainable parameter, initialized at $c_m = 0$ (that is $M = \sigma$), so that only $\sigma$ and $J$ are prescribed and the mass is a prediction through (\ref{eq:M_pred}).

For the nearly extremal spin $a/M = 0.99$ with fixed mass, where $c_m \approx 12.2$ is large, the first stage stalls when started from random weights. In this case we start from the network obtained in the free-mass mode for the same spin and continue with the second stage only, now with the mass fixed. The free-mass network itself is trained without any information about the solution.

Finally, we report the networks at the end of the training. In particular, we do not select the best intermediate state, since this would require knowledge of the exact solution, which enters only in the evaluation of the results in Sec.~\ref{sec:results}. We trained $84$ networks: in each of the two modes, ten networks with different random initializations for $a/M = 0.5$, $0.95$ and $0.99$, and three for $a/M = 0.1$, $0.3$, $0.7$ and $0.9$. A single training takes about a quarter of an hour on one CPU core.
% Claude draft, 2026-10-26

\section{Results}
\label{sec:results}
% ---------------------------------------------------------------------------
% Sec. "Results". Drafted by Claude (AI) on 2026-10-26 in the style of the
% author's sections; to be edited by the author. Numbers: terminal networks of
% the c-18 runs, compared with exact Kerr after training by
% prolate_nn/paper_metrics.py (figures/paper_metrics.json); figures by
% prolate_nn/plot_panels.py, plot_spin_maps.py and plot_trajectories.py.
% ---------------------------------------------------------------------------

In this section we compare the trained networks with the exact Kerr solution (\ref{eq:kerr})--(\ref{eq:kerr_class}) for $M = 1$ and spins $a/M = 0.1$, $0.3$, $0.5$, $0.7$, $0.9$, $0.95$ and $0.99$. Since the networks represent the fields on the whole compactified rectangle, we can compare them with the exact solution everywhere, including the horizon, the axis and spatial infinity. We use the maximal relative error of $F$ over the rectangle, $\epsilon_F = \max|F_{\rm NN}/F_{\rm exact} - 1|$, its average $\langle\epsilon_F\rangle$ over the rectangle, and the maximal error of the twist potential normalized by its value on the axis, $\epsilon_B = \max|B_{\rm NN} - B_{\rm exact}|/4J$. Note that $F_{\rm NN}/F_{\rm exact} = e^{U_{\rm NN} - U_{\rm exact}}$, therefore the relative error of $F$ is essentially the absolute error of $U$, and it remains finite on the axis and on the horizon. We also follow how far the action is above its Kerr value, $(E - E_{\rm Kerr})/E_{\rm Kerr}$, with both evaluated by the same quadrature.

\begin{table}[t]
\caption{Accuracy of the trained networks for all spins. The first three columns refer to the fixed-mass mode, and the last one to the free-mass mode, in which the mass is predicted. Each entry is the median over the trained networks: ten for $a/M = 0.5$, $0.95$ and $0.99$, and three otherwise.}
\label{tab:spins}
\begin{ruledtabular}
\begin{tabular}{ccccc}
$a/M$ & $\epsilon_F$ & $\langle\epsilon_F\rangle$ & $\epsilon_B$ & $|M_{\rm pred} - M|/M$ \\
\hline
0.1 & $2.1\times10^{-5}$ & $2.1\times10^{-6}$ & $1.4\times10^{-5}$ & $5.2\times10^{-6}$ \\
0.3 & $6.0\times10^{-5}$ & $1.0\times10^{-5}$ & $7.9\times10^{-5}$ & $1.8\times10^{-4}$ \\
0.5 & $1.4\times10^{-4}$ & $1.4\times10^{-5}$ & $1.1\times10^{-4}$ & $5.0\times10^{-4}$ \\
0.7 & $1.6\times10^{-4}$ & $1.4\times10^{-5}$ & $7.4\times10^{-5}$ & $5.0\times10^{-4}$ \\
0.9 & $2.8\times10^{-4}$ & $1.2\times10^{-5}$ & $1.1\times10^{-4}$ & $3.5\times10^{-4}$ \\
0.95 & $3.2\times10^{-4}$ & $2.4\times10^{-5}$ & $1.4\times10^{-4}$ & $6.9\times10^{-4}$ \\
0.99 & $7.1\times10^{-4}$ & $5.1\times10^{-5}$ & $3.2\times10^{-4}$ & $1.7\times10^{-3}$ \\
\end{tabular}
\end{ruledtabular}
\end{table}

\subsection{The recovered fields}
\label{sec:fields}

\begin{figure*}[t]
\centering
\includegraphics[width=\textwidth]{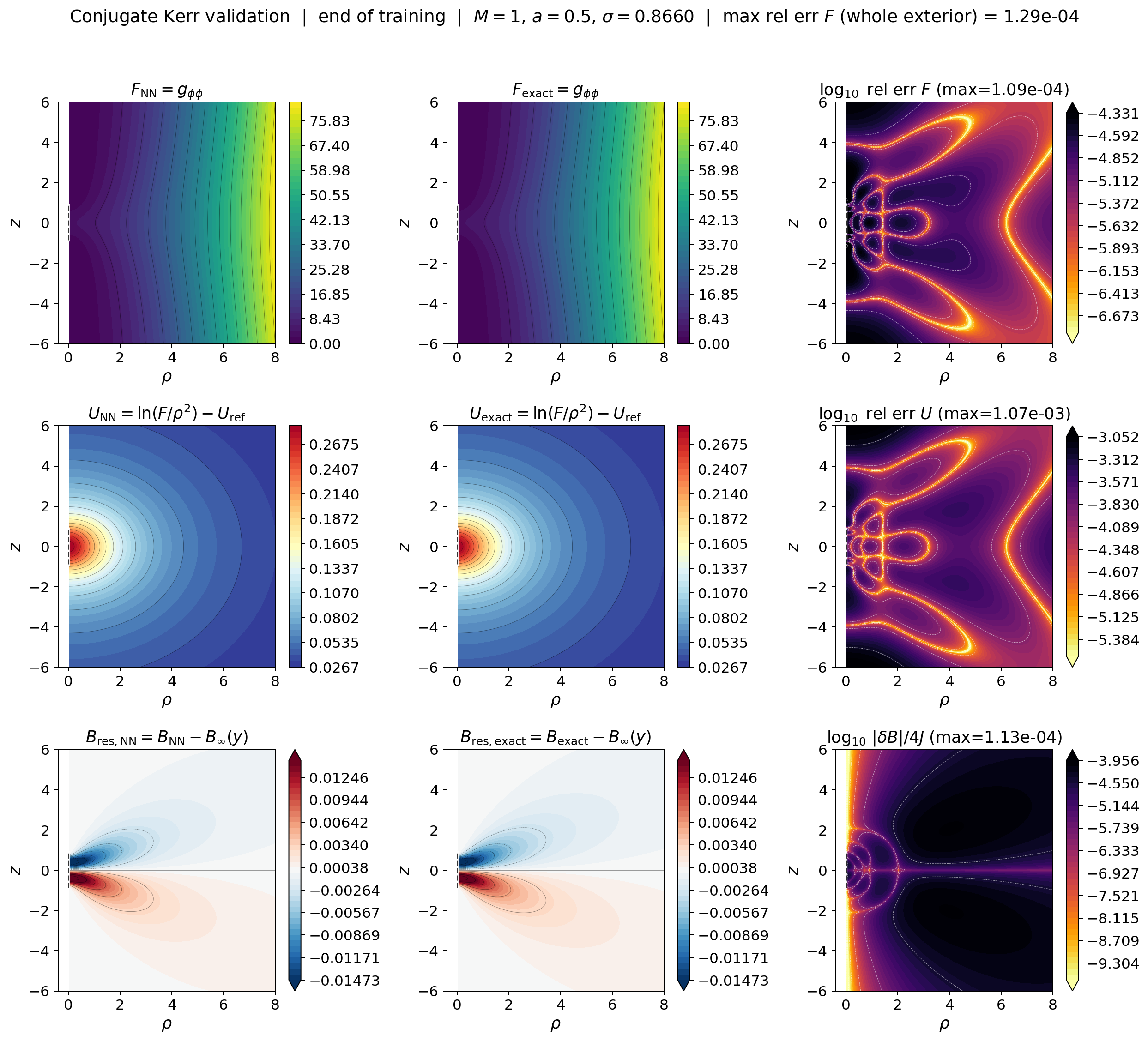}
\caption{Recovered $F$, $U$ and $B$ for $a/M = 0.5$ with fixed mass, at the end of the training. Top row: $F_{\rm NN}$, $F_{\rm exact}$ and the $\log_{10}$ relative error in $F$. Middle row: $U_{\rm NN}$ vs.\ $U_{\rm exact}$, the field learned by the network in (\ref{eq:F_split}), and the $\log_{10}$ relative error in $U$. Bottom row: the part of the twist potential that is not fixed by the boundary data, $B - B_\infty(y)$, for the network and the exact solution, together with the $\log_{10}$ error in $B$ normalized by $4J$. The horizon rod is marked in white.}
\label{fig:panels}
\end{figure*}

Let us start with $a/M = 0.5$. In Fig.~\ref{fig:panels} we show the recovered fields with fixed mass for one of the ten trained networks, chosen as the median by $\epsilon_F$. In each row the network and the exact solution are indistinguishable by eye, and the third column shows the errors. The relative error of $F$ is below $1.1\times10^{-4}$ in the plotted region, and $\epsilon_F = 1.3\times10^{-4}$ over the whole exterior, reached at the ends of the horizon rod, where the horizon meets the axis. The bright lines in the error maps are the curves on which the error changes sign.

The relative error of $U$ is about ten times larger, since $U$ itself is small: it varies between $0.03$ and $0.29$ in the plotted region. The error of $B$ is at most $1.1\times10^{-4}$ in units of $4J$, and it vanishes on the axis, where $B$ is fixed by the boundary conditions. Note that most of the twist potential is fixed by its asymptotic form (\ref{eq:Binf}): the part learned by the network, $B - B_\infty$, is about one percent of $4J$, and the network recovers it with a relative accuracy of about one percent.

\subsection{Dependence on the spin and the predicted mass}
\label{sec:spins}

% Figs. spins and traj share one float, so that they appear on the same page before the appendix.
\begin{figure*}[tp]
\centering
\includegraphics[width=\textwidth]{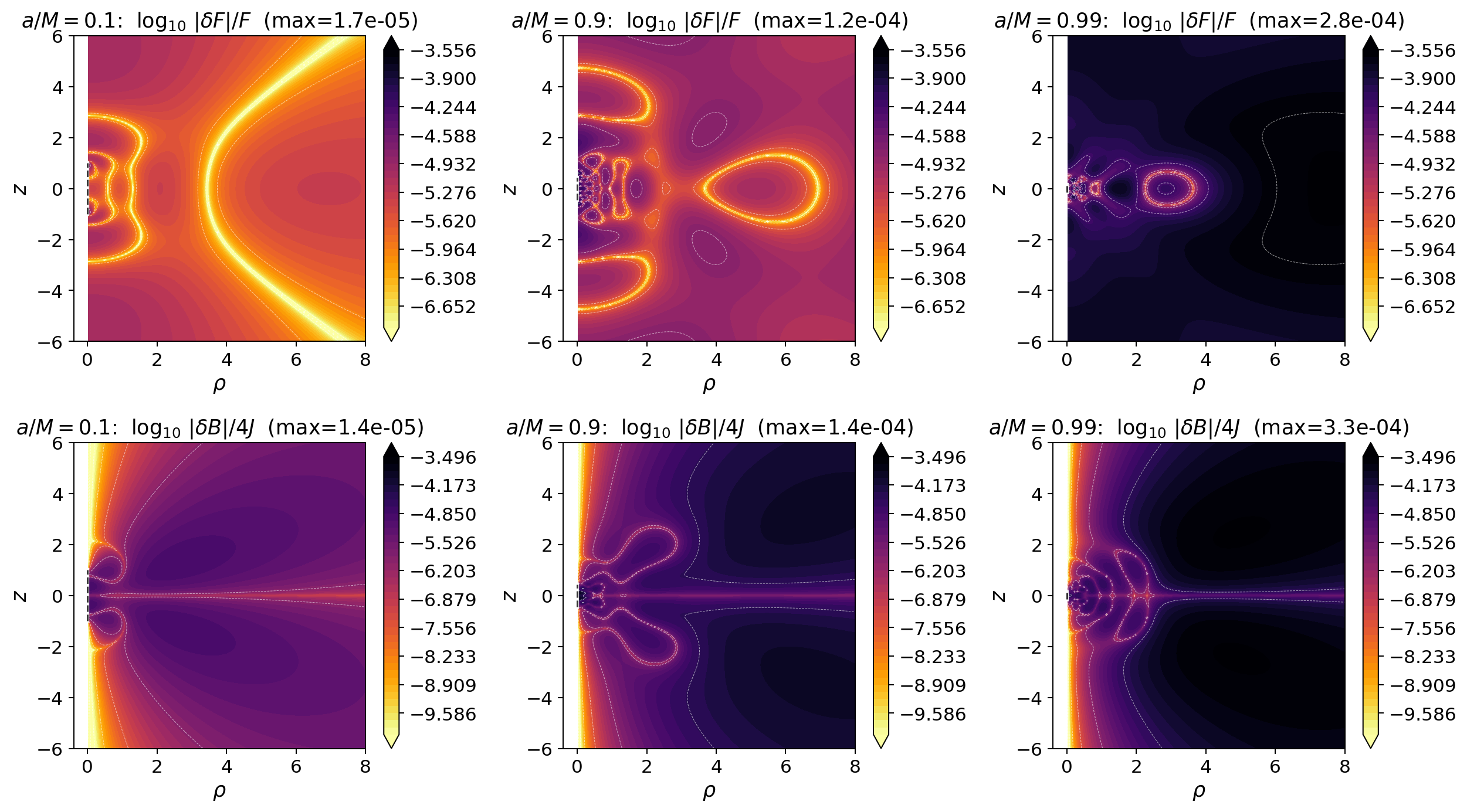}
\caption{Errors of the recovered $F$ (top) and $B$ (bottom) for $a/M = 0.1$, $0.9$ and $0.99$ with fixed mass, on a common color scale in each row. These are the analogues of the third column of Fig.~\ref{fig:panels}; the error of $U$ is not shown, since it follows the error of $F$. For $a/M = 0.99$ the training starts from the free-mass network (see Sec.~\ref{sec:modes}).}
\label{fig:spins}
\vspace{0.5em}
\centering
\includegraphics[width=0.72\textwidth]{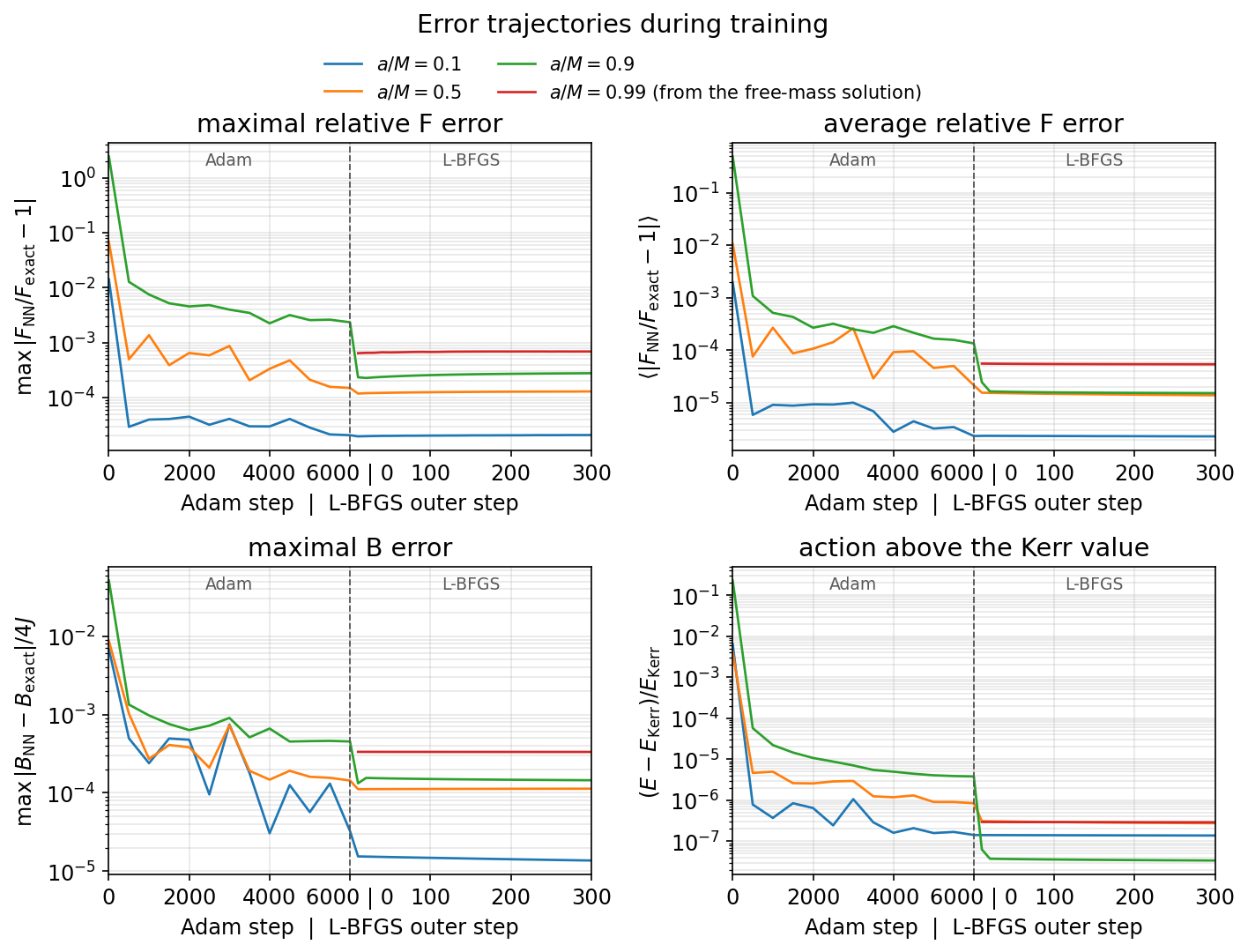}
\caption{Error measures during the training of the median networks for $a/M = 0.1$, $0.5$, $0.9$ and $0.99$ with fixed mass: the maximal and the average relative error of $F$ over the whole exterior (top), the maximal error of $B$ in units of $4J$ and the relative excess of the action over its Kerr value (bottom). Left part of each panel: the Adam stage; right part: the L-BFGS stage. The network for $a/M = 0.99$ starts from the free-mass solution.}
\label{fig:traj}
\end{figure*}

In Fig.~\ref{fig:spins} we show the errors of $F$ and $B$ for $a/M = 0.1$, $0.9$ and $0.99$ on a common color scale. The accuracy decreases as the spin approaches its extremal value, while the structure of the errors is similar for all spins. In all cases the largest error of $F$ is found at the ends of the horizon rod, that is, at the corners $(s,y) = (1,\pm1)$ of the rectangle. Table~\ref{tab:spins} summarizes the results for all spins: $\epsilon_F$ grows from $2\times10^{-5}$ at $a/M = 0.1$ to $7\times10^{-4}$ at $a/M = 0.99$, while the average error $\langle\epsilon_F\rangle$ stays below $5\times10^{-5}$.

In the free-mass mode the networks predict the mass of the black hole through (\ref{eq:M_pred}). As can be seen from the last column of Table~\ref{tab:spins}, the prediction is accurate to $5\times10^{-6}$ for $a/M = 0.1$ and to $2\times10^{-3}$ for $a/M = 0.99$. Interestingly, for $a/M \geq 0.3$ the predicted mass is slightly below the true one in all runs. The accuracy of the fields in this mode is comparable to that of the fixed-mass mode.

\subsection{Training dynamics}
\label{sec:dynamics}

Finally, let us look at how the networks approach the solution during the training. In Fig.~\ref{fig:traj} we show the error measures and the action for the median networks at $a/M = 0.1$, $0.5$, $0.9$ and $0.99$. The first stage reduces the errors by two to three orders of magnitude, most of it within the first $500$ steps. The L-BFGS stage then gives a further improvement, mostly within its first ten steps, after which the field errors change little while the action continues to decrease slowly. In all cases the action approaches its Kerr value from above, as expected for a minimum. Since the Kerr solution minimizes the action, the gap $E - E_{\rm Kerr}$ is quadratic in the field errors, which is why a relative gap of order $10^{-7}$ goes together with field errors of order $10^{-4}$.

The remaining error depends only weakly on the details of the training. At $a/M = 0.5$, wider or deeper networks (up to $64$ neurons or four hidden layers) and a three times longer first stage reduce $\epsilon_F$ by up to about a factor of two, while a finer training rule does not improve it. We have not attempted to push the accuracy further.
% Claude draft, 2026-10-26

\begin{acknowledgments}
This work was supported by the Bulgarian NSF grant KP-06-N88/1.
The partial support of KP-06-N62/6 from the Bulgarian science fund
is also gratefully acknowledged.
\end{acknowledgments}

\appendix

\section{Regularity and boundary terms of the admissible class}
\label{app:class}
% Appendix to Sec. sec:class, drafted by Claude (AI) on 2026-10-18; moved here
% from the main text at the author's request.
In this appendix we show that, for fields in the admissible class (\ref{eq:U_class})--(\ref{eq:B_class}), the integrand of (\ref{eq:E_rect}) is bounded on the closed rectangle and all boundary terms in its first variation vanish.

Let us start with the integrand. Using (\ref{eq:U_class})--(\ref{eq:B_class}) and $F = \sigma^2(1+s)^2(1-y^2)\,e^{U}/s^2$, we have:
\begin{equation}
\begin{split}
    U_y &= s^2 N_y , \qquad
    \frac{1}{F^2} = \frac{s^4\, e^{-2U}}{\sigma^4(1+s)^4(1-y^2)^2} , \\
    B_y &= (1-y^2)\bigl[-6J + s^2\bigl((1-y^2)\,G_y - 4y\,G\bigr)\bigr] , \\
    B_s &= (1-y^2)^2\bigl(2s\,G + s^2 G_s\bigr) .
\end{split}
    \label{eq:class_derivs}
\end{equation}
Therefore, the factor $1/s^2$ in the $U$ term of (\ref{eq:E_rect}) is canceled by $U_y^2 \propto s^4$, while in the $B$ terms the factors $1/s^2$ are canceled by $F^{-2} \propto s^4$. Note that the factor $(1-y^2)^{-2}$ in $F^{-2}$ is canceled by $B_y^2 \propto (1-y^2)^2$ and $B_s^2 \propto (1-y^2)^4$. As a result, the integrand of (\ref{eq:E_rect}) is bounded everywhere on the closed rectangle, including the horizon $s = 1$, the axis $y = \pm1$ and spatial infinity $s = 0$.

Next, we consider the first variation of (\ref{eq:E_rect}). Since $F$ enters without derivatives, only the gradient terms produce boundary contributions after integration by parts:
\begin{equation}
\begin{split}
    \delta E\big|_{\partial} = 2\sigma\!\int_{-1}^{1}\!\!dy\,
      &\Bigl[(1-s^2)\Bigl(U_s\,\delta U + \frac{B_s\,\delta B}{F^2}\Bigr)\Bigr]_{s=0}^{s=1} \\
    + 2\sigma\!\int_0^1\!\!ds\,
      &\Bigl[\frac{1-y^2}{s^2}\Bigl(U_y\,\delta U + \frac{B_y\,\delta B}{F^2}\Bigr)\Bigr]_{y=-1}^{y=1} .
\end{split}
    \label{eq:bdy_terms}
\end{equation}
On the horizon ($s = 1$) and on the axis ($y = \pm1$) these terms vanish due to the explicit factors $1-s^2$ and $1-y^2$, since by (\ref{eq:class_derivs}) the remaining factors are bounded there. At spatial infinity ($s = 0$) they vanish because the admissible variations, $\delta U = s^2\delta N$ and $\delta B = (1-y^2)^2 s^2\delta G$, vanish there, while $U_s$ and $B_s/F^2$ remain finite. The same holds for the variation $\delta U = s\,\delta c_m$ when $c_m$ is treated as unknown. Therefore, the natural boundary conditions on the horizon and on the axis are satisfied identically, and no boundary terms need to be added to (\ref{eq:E_rect}).

\bibliography{references}
\end{document}